\documentclass[pre,twocolumn,amsmath,amssymb,superscriptaddress]{revtex4-2}
\usepackage{graphicx}
\usepackage{color,soul}
\usepackage[caption=false]{subfig}
\usepackage{amsmath}
\usepackage{float}
\usepackage[normalem]{ulem}
\usepackage{xr-hyper}
\usepackage{romannum}
\usepackage{gensymb}
\usepackage{makecell}
\usepackage{setspace}
\usepackage{placeins}

\usepackage[colorlinks=true,linkcolor=blue,anchorcolor=black,
citecolor=magenta,filecolor=blue,menucolor=black,
runcolor=black,urlcolor=blue]{hyperref}

\newcommand{\SK}[1]{\textcolor{black}{{#1}}}

\newcommand{\SD}[1]{\textcolor{red}{{#1}}}
\begin{document}
\pagenumbering{arabic}
\title{Anomalous Dynamical Heterogeneity in Active Glasses as a Signature of Violation of Mermin-Wagner-Hohenberg Theorem}
\author{Subhodeep Dey}
\affiliation{Tata Institute of Fundamental Research Hyderabad, 36/P, Gopanpally Village, Serilingampally Mandal, Ranga Reddy District,
Hyderabad, Telangana 500046, India}
\author{Smarajit Karmakar}\email{smarajit@tifrh.res.in}
\affiliation{Tata Institute of Fundamental Research Hyderabad, 36/P, Gopanpally Village, Serilingampally Mandal, Ranga Reddy District,
Hyderabad, Telangana 500046, India}
\begin{abstract}
\SK{Two-dimensional (2D) systems have attracted renewed interest within the scientific community due to their anomalous dynamical behaviors, which arise from long-wavelength density fluctuations as predicted by the Mermin-Wagner-Hohenberg (MWH) theorem. In equilibrium, it is well established that continuous spontaneous symmetry breaking (SSB) in 2D is prohibited at any finite temperature ($T > 0$), resulting in the absence of true long-range positional order and establishing $d_l = 2$ as the lower critical dimension. Recent studies have demonstrated that, in active systems, the lower critical dimension can shift from $d_l = 2$ to $3$. This study examines the impact of MWH theorem violation in active systems on dynamical heterogeneity (DH). As a minimal model, glassy systems of active particles undergoing run-and-tumble (RT) motion are considered. Glass-like dynamical behavior, including anomalously enhanced DH, is observed in various biological systems such as collective cell migration, bacterial cytoplasm, and ant colonies. Furthermore, the study investigates the influence of local positional order, or medium-range crystalline order (MRCO), on DH in the presence of activity. The results indicate that the growth of DH with increasing activity differs significantly between systems with and without MRCO. These findings may have important implications, as many biological systems exhibit local structural ordering, and DH could serve as a useful indicator for quantifying the degree of ordering.}
\end{abstract}

\maketitle
\noindent{\bf \large Introduction:}
\SK{The glassy systems are intriguing because of their rapid increase in relaxation time and viscosity, despite being structurally disordered. Such systems are commonly found in nature, spanning from microscopic to macroscopic scales, and existing in both living and non-living entities. There are unique dynamical properties associated with glassy systems, including the caging effect on particles caused by their neighbors, a two-step relaxation process, and dynamic heterogeneity (DH). Two-dimensional (2D) systems, on the other hand, are noteworthy due to the presence of long-wavelength density fluctuations as a consequence of the Mermin-Wagner-Hohenberg (MWH) theorem, which predicts that in equilibrium, establishment of long-range positional order at any finite temperature ($T > 0$) is not possible for a particulate system with continuous degrees of freedom. Basically, this means that the breakdown of continuous spontaneous symmetry in 2D is not possible in equilibrium at any finite temperature. This phenomenon has been demonstrated in a series of studies involving spin and particle systems \cite{Mermin1966, Mermin1968, Hohenberg1967}. Using computer simulations, Shiba et al. \cite{Shiba2016} demonstrated that disordered  systems in 2D exhibit fluctuations that diverge with increasing system size, as predicted by the MWH theory as well. At the same time, experimental studies on 2D colloidal systems have confirmed the presence of the MWH fluctuations \cite{Flenner2015, Vivek2017, Illing2017}.
}

\SK{It is natural to ask whether the same physics applies to non-equilibrium situations especially in their steady states at which various effective equilibrium like pictures (e.g. effective temperature etc.) are found to describe the physics quite well. Some ubiquitous non-equilibrium systems of significant interests both for fundamental sciences and day-to-day applications are active matters. When constituents of a matter move under the influence of its internal energy in addition to the thermal fluctuations present in the system due to a heat bath, then one often classified these systems as active matters \cite{Vicsek1995, Toner1998, Ramaswamy2010, Palacci2013, Marchetti2013}. Thus, active matters will inherently be in out of equilibrium states and will not follow detailed balance conditions essential for thermal equilibrium. Due to the presence of collective motion, active matter can exhibit true long-range order even in 2D \cite{Vicsek1995, Toner1998} in stark contrast with the Mermin-Wagner-Hohenberg (MWH) theory for the equilibrium system. Recently, it has been shown that an assembly of particles in the presence of active forcing can lead to the formation of a 2D ideal crystal at high enough packing fraction\cite{Galliano2023}. The plethora of non-trivial behaviour observed in systems with various non-equilibrium driving led to the emergence of yet another research direction in the area of dense disordered systems known as active glasses \cite{Mandal2016, Paul2023, Dey2022}. Glass-like dynamical behaviour has been found in many biological systems such as epithelial cell monolayer tissue, cancerous cell proliferation, and wound-healing processes \cite{Zhou2009, Angelini2011, Parry2014, Park2015, Garcia2015, Malinverno2017, Nishizawa2017, Kim2020, Vishwakarma2020, Cerbino2021, Sadhukhan2024}. Again, modulating the presence of ATP in the cell cytoplasm and bacterial cytoplasm, the glassy dynamics of the system can be tuned \cite{Zhou2009, Parry2014}.
}

\SK{
Naturally, these observations lead to the emerging curiosity of how activity might alter the physics of glass transition in these systems and whether glassy dynamics, as regulated by activity, play any important role in biological functions. In a recent work \cite{Sharma2025}, it has been shown that active driving can lead to better annealing of these systems in a manner that has a strong parallel with the annealing due to oscillatory shear. It has also been shown that activity-induced annealing can help a material to change its rigidity significantly, and thus active annealing might be an important tool in the maturation of bone tissues. Another recent work \cite{Wei2023} on active gel demonstrated that annealing achieved via light-sensitive active particles for half an hour is equivalent to thermal annealing for nearly three months. Thus, a detailed investigation of the dynamical behaviour of active glassy systems might open up new possibilities for both material design and applications in the near future.
}

\SK{
Although active systems are in out-of-equilibrium conditions, there have been significant attempts to see if the statistical properties of these systems remain equilibrium-like in steady state conditions at a small activity limit. In some works, it has been shown that some of the dynamical aspects of the system can be well-described by an effective temperature via a generalized fluctuation-dissipation theory (FDT). In \cite{Loi2008}, it has been shown that an effective temperature can be defined via an FDT analysis. Again, for the active Ornstein-Uhlenbeck process, a generalized FDT can be defined using the effective potential, and the dynamics follow the time-reversal symmetry \cite{Fodor2016} at a small activity limit. This effective temperature-like description of the dynamics of active glasses is also explored recently, where the effective temperature is found to rationalize some of the dynamical properties of the system \cite{Nandi2018, Berthier2013, Berthier2014, Ni2013, Mandal2016, Nandi2017, Berthier2019}. However, it has been shown in \cite{Paul2023} that the effective temperature description is not useful for all dynamical quantities. In particular, the dynamic heterogeneity (DH) of the active glass can not be explained by using the effective temperature description.
}

\SK{
Dynamic heterogeneity (DH) is an important hallmark of the glassy dynamics. It highlights the existence of differently relaxing regions in the system, with some relaxing much faster and some slower than the average degree of relaxation in the system \cite{Cavagna2009, Berthier2011, Karmakar2014}. One of the well-accepted methods of quantifying the DH is the measurement of four-point dynamic susceptibility ($\chi_4(t)$, see definition in SM). In a recent work of an active glassy system in 3D \cite{Dey2022}, it has been shown that the presence of active driving leads to an emergence of an additional short-time peak in $\chi_4(t)$, which enhances with increasing strength of activity. Previous studies have shown the existence and emergence of such a peak  in systems at equilibrium when the linear size of the system is increased. It was clearly demonstrated that the peak is purely due to the long-wavelength phonon modes in the system, and if one damps these modes, then the peak disappears. Thus, enhancement of this short-time peak in $\chi_4(t)$ with increasing activity signals the enhancement of these modes due to efficient coupling of the low-frequency modes with the active driving in a non-trivial manner. In \cite{Dey2025}, it was highlighted that the MWH theorem is violated in active solids in both 2D and 3D, especially in 2D, activity enhances long-wavelength density fluctuations to such an extent that the Debye-Waller (DW) factor exhibits power-law divergence with system size much stronger than the well-known logarithmic divergence in 2D solids in equilibrium. Just to emphasise the degree of enhancement of these long-wavelength phonon modes due to activity, it has demonstrated that active solids, even in 3D, are unstable to these fluctuations as the DW factor exhibited logarithmic divergence with system size in 3D. Only in four-dimensions the active solids under these active driving, found to be thermodynamically stable. This suggests an overall shift in the lower critical dimensions $d_l = 2$ in equilibrium systems to $d_l = 3$ in active systems. Thus, enhanced phonon activity at long wavelength is expected to have a very strong effect on the DH in active glassy systems in 2D as well as in 3D.
}

\SK{
In this article, we extensively study the effect of activity on the dynamics of two glass-forming model systems in 2D. These glassy models, referred to as 2dmKA and 2dKA, are characterized by varying degrees of local structural ordering. The former is a generic model of glass-forming liquids, while the latter exhibits growing local or medium-range crystalline order (MRCO) as temperature decreases. These models are described in detail in the Methods Section.  We have also identified a unique feature of the effect of the activity in two different model glass-formers; specifically, we have observed that in the 2dmKA model, the long-time peak of $\chi_4(t)$ decreases with increasing activity, along with a decrease in the characteristic timescale. In contrast, for the 2dKA model with MRCO, we have seen that the peak height of $\chi_4(t)$ increases with increasing activity, although the characteristic timescale decreases. Our study shed light on the surprising and unique feature of systems with local or medium-range crystalline order (MRCO). This may explain the recent experiments on cellular monolayers, where it was discovered that dynamic heterogeneity (DH) increases with increasing activity in the medium, while the relaxation time decreases systematically \SD{\cite{Cerbino2021}}. Additionally, we have performed Brownian dynamics simulations of these models to compare and contrast their behavior with that of the active systems. Further details of the models and simulation protocols are provided in the Method section.
}

\begin{figure*}[!htb] 
	\includegraphics[width=1.0\linewidth]{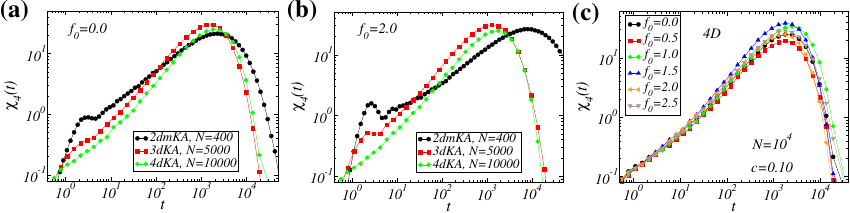}
	\caption{{\bf MWH Fluctuations across Dimensions:} \SK{(A), (B) and (C) show $\chi_4(t)$ as a function of time for different dimensions and various activities.  (A) Shows the dimensional effect on dynamical heterogeneity for the passive system in 2, 3, and 4 dimensions; the system size (L) is of the same order across dimensions. For a 2D passive system, there is a pronounced short-time peak of $\chi_4(t)$. Similarly, (B) shows the dimensional effect for the active system ($f_0=2.0$) for 2, 3, and 4 dimensions. Notice the short-time peak gets enhanced in the presence of activity in 2D and a appearance of a peak in 3D system. (C) $\chi_4(t)$ for different activities for 4D system. Note that there is no short time peak in $\chi_4(t)$, for system size $N=10000$ in 4D. The short-time peak of $\chi_4(t)$ gets enhanced for both 2D and 3D, but in the 4D system, it is not present. This is in complete agreement with the shifting of the lower critical dimension of the system to $d_l=3$ \cite{Dey2025}.}}
	\label{fig:chi4_MWH_Plots}
\end{figure*}
\vskip +0.2in
\noindent{\bf \large Methods Section: }\\
{\bf Model \& Simulation Details: }
\SK{We have studied the dynamics of two model glass-forming liquids. The first glass-forming liquid model is the 2dmKA Binary model with a number ratio of $65:35$ (large : small) with the other details of the potential same as that of the Kob-Anderson model \cite{Das2017} (see SM for details). This particular number ratio ensures that there are no tendencies for the system to form local crystalline orders. In contrast, the other model is referred to as a 2dKA binary mixture with $80:20$ (large : small) particle number ratio, and it forms local medium range crystalline order (MRCO) at low temperatures \cite{Kawasaki2007, Tanaka2010,  Tah2018}. We have performed a simulation with a number of particles, $N \in [400, 50000]$. We ran $32$ statistically independent ensembles for all the systems except the few large ones ($25000 - 50000$); we have taken $8$ ensembles for these system sizes. We used a three-chain No\'se-Hoover thermostat to perform NVT simulations \cite{Martyna1992}. For 4D Kob-Andersen model (4dKA), A:B number ratio is $80:20$ with density ($\rho$) is $1.474$ \cite{Eaves2009}.
}

\noindent{\bf Modelling Activity: }
\SK{The activity in the system is introduced in the form of run and tumble particle (RTP) dynamics \cite{Mandal2016, Paul2023, Dey2022}, where the dynamics of the active particles can be tuned using three parameters such as concentration of active particles, $c$, strength of the active force, $f_0$, and persistence time of the active motion, $\tau_p$. For setting up the reference temperature, we kept $\tau_p = 1$, active force magnitude $f_0$ is selected from $0.0 - 2.0$ for 2dmKA and $0.0 - 2.8$ for 2dKA by fixing concentration $c=0.1$. When we varied the concentration $c\in [0.0, 0.6]$, we kept active force $f_0 = 1.0$ for both the 2dmKA and 2dKA systems. We integrate the following modified equations of motion in the presence of active driving,
\begin{eqnarray}
\dot{\vec{r}}_i&=&\frac{\vec{p}_i}{m}\nonumber\\
\dot{\vec{p}}_i &=& -\frac{\partial \Phi}{\partial \vec{r}_i} + \eta_i\vec{F}_i^A,
\end{eqnarray}
where $\vec{r}_i$ and $\vec{p}_i$ are the position and momentum vector of $i^{th}$ particle, $\eta_i$ is the activity-tag which take values $1$ or $0$ depending on whether the particle is active or passive, $\Phi$ is the inter-particle potential, and $\vec{F}_i^A$ is the active force on $i^{th}$ particle.  In 2D, the active force can be written as
\begin{equation}
    \vec{F}_i^A = f_0(k_x^i \hat{x} + k_y^i \hat{y}),
\end{equation}
where ($k_x,k_y$) are chosen from $\pm 1$, such that $\sum_{\alpha,i} k_{\alpha}^i=0$, i.e., the net active momentum along any direction is zero. This can be easily generalised to higher dimensions. The run-and-tumble motions of active particles keep the inertial effect of the system preserved; this will not be the case for the ABP (Active Brownian particle) model. Recent work on active matter suggests that the inertial term is important to understand the system \cite{teVrugt2023}.
}

\vskip +0.2in
\noindent{\bf \large Results: }
\begin{figure*}[!htb] 
	\includegraphics[width=1.0\linewidth]{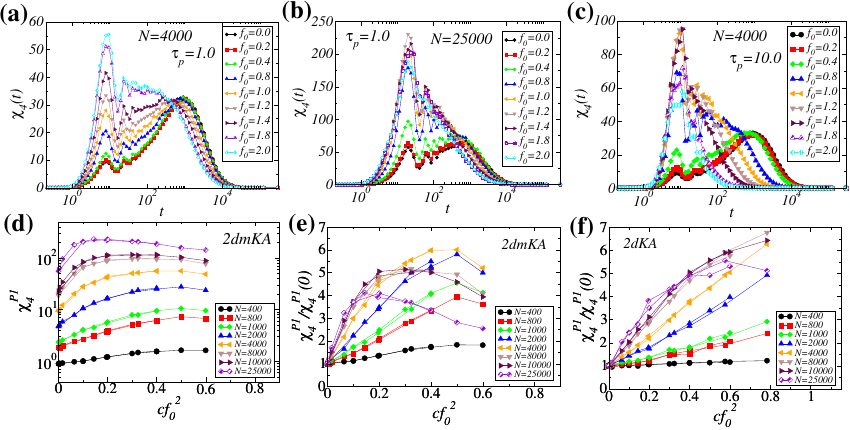}
	\caption{{\bf Dynamic Heterogeneity:}
 (a), (b) and (c) shows $\chi_4(t)$ as function of time for 2dmKA system. (a) shows steady increase in first peak of $\chi_4(t)$ masking the $\chi_4(t)$ peak around $\tau_{\alpha}$ for system size $N = 4000$ with changing activity $f_0$, fixing $c = 0.1$ and $\tau_p = 1.0$, 2dmKA. (b) Same with increased system size $N = 25000$, keeping all parameters the same as (a), we can observe a drastic increase in the first $\chi_4$ peak. (c) Results with increased persistent time $\tau_p=10$, for system size $N=4000$. (d) $\chi_4^{P1}$ as function of effective activity $cf_0^2$ for 2dmKA system. (e) Scaled $\chi_4^{P1}$ with increasing activity with respect to the passive system. For the 2dmKA system $\chi_4^{P1}$/$\chi_4^{P1}(0)$ shows saturation at some characteristic effective activity, which tends to decrease with increasing system size, with a broader peak. (f) For the 2dKA system $\chi_4^{P1}$/$\chi_4^{P1}(0)$ does not show a rapid decrease in characteristic activity with increasing system size like (e).
}
	\label{fig:chi4_Plots}
\end{figure*}
\vskip +0.05in
\noindent{\bf MWH Fluctuations Across Dimensions:} 
\SK{To highlight the effect of dimensionality, primarily coming due to enhanced long wavelength fluctuations in active systems, on the dynamical heterogeneity, we show the four-point dynamical susceptibility, $\chi_4(t)$ in Fig.\ref{fig:chi4_MWH_Plots} as a function of time for different activities and across dimensions. In Fig.\ref{fig:chi4_MWH_Plots} (A), a pronounced short-time peak due to the long-wavelength mode can be observed for 2D passive system. In the presence of activity, this peak is found to be enhanced strongly in 2D and appearance of a peak is observed in 3D as shown in Fig. \ref{fig:chi4_MWH_Plots} (B). $\chi_4(t)$ is computed for the active systems at $f_0=2.0$ in 2D, 3D and 4D respectively. Note that we have take systems sizes across dimensions in such a manner that the linear dimension of the box is nearly the same for a better comparison. If we choose a much bigger system size sizes then one would expect the short-time peak to be much bigger in both 2D and 3D. This short time peak in 3D has already been shown to grow systematically to larger values with increasing activity even if the system size is kept fixed. In Fig.\ref{fig:chi4_MWH_Plots}(C), we show $\chi_4(t)$ in 4D and one can clearly see that there is no short-time peak in $\chi_4(t)$, for a system size $N=10000$ in for all the different activities studied. The appearance of a short-time peak and enhancement of the same with increasing activity clearly suggest that low-frequency long-wavelength phonon modes are enhanced in active systems in both 2D and 3D, and are not in 4D. This occurrence of the anomalous dynamic heterogeneity in 2D and 3D systems is in complete agreement with the shifting of the lower critical dimension of the system to $d_l=3$ \cite{Dey2025}, which signifies a strong departure from the predictions of Mermin-Wagner-Hohenberg (MWH) theorem in equilibrium.}

\vskip +0.05in
\noindent{\bf Effect on Dynamic Heterogeneity:}
\SK{In this section, we focus on the influence of long wavelength phonon modes on the dynamical heterogeneity (DH), as quantified by the four-point susceptibility, $\chi_4(t)$ \cite{Dey2022}. In Fig.\ref{fig:chi4_Plots}(A), we show $\chi_4(t)$ for $N = 4000$ for varying active force amplitude, $f_0$ at a constant persistence time, $\tau_p = 1.0$. One can clearly see that short-time peak systematically increases with increasing the activity and for $f_0 = 2.0$, the short-time peak is much larger than the peak at $\alpha$-relaxation time. In fact for large activities, the long time peak nearly cease to exist. As this peak is solely due to long wave length fluctuations, so one expects that the effect of activity will be much stronger for larger system sizes. In Fig.\ref{fig:chi4_Plots}(B), we show the similar data but for much bigger system size of $N = 25000$ particles. One can see that the short-time peak reached a value close to $250$ which is nearly four times larger than $N=4000$ system size result. For $N=25000$, due to immense increase in the short-time peak height, the peak at $\alpha$-relaxation time is not visible. In Fig.\ref{fig:chi4_Plots}(C), we show again similar results but for $\tau_p = 10$ and system size $N=4000$. If one compares the results of panel (A) and (C) then it can be clearly seen that increase in persistence time can lead to strong enhancement of the long-wavelength phonon modes as evident from the rapid increase of the short-time $\chi_4(t)$ peak. Our findings thus strongly indicate that increasing activity leads to the appearance and enhancement of a short-time peak in $\chi_4(t)$, and the effect is much pronounced in two dimensions compared to three dimensions. Additionally, we observe the appearance of additional peaks or oscillations in $\chi_4(t)$, which are believed to be related to the higher harmonics of the frequencies. Interestingly, we find that the first peak of $\chi_4(t)$ diverges with system size ($L$), as reported for the 2D passive glass in previous studies \cite{Shiba2016}. However, the enhancement of this peak with increasing activity has not been reported before. Our results raise important questions about the measurement of DH in various experiments done in two dimensions with different active drivings, such as external vibration for granular medium, ATP-driven activity in biological systems, or chemical driving in chemophoretic particles. Even for Janus colloids, the effect will be much stronger due to the activity \cite{Li2019}.
}

\SK{Fig.\ref{fig:chi4_Plots}(d) show the first peak height of $\chi_4(t)$, referred to as $\chi_4^{P1}$, for 2dmKA model as a function of activity parameter $cf0^2\tau_p$ (here $\tau_p =1$) which can be shown to be a good parameter to uniquely quantify the degree of activity in the system in the small activity limit. We show the results for various system sizes, one can see that for large enough system sizes, $\chi_4^{P1}$ first increases with increasing activity but then it starts to decrease giving rise to a broad peak at an intermediate activity. In  Fig.\ref{fig:chi4_Plots}(e) we show the same data but now rescaled by the value at the zero activity to highlight the non-monotonic dependence.  One also sees that the relative increase of $\chi_4^{P1}$ with increasing activity actually starts to become weaker as one increases system size for the 2dmKA model.  Results for the 2dKA model as shown in Fig.\ref{fig:chi4_Plots}(f) are very different. One does not see a peak like feature for 2dKA model although for large system size there is some signature of it but at very large activity. This interesting difference leads us to explore the possible effect of local ordering on these dynamical aspects of the systems.
}

\begin{figure}[!htb] 
	\includegraphics[width=1.0\linewidth]{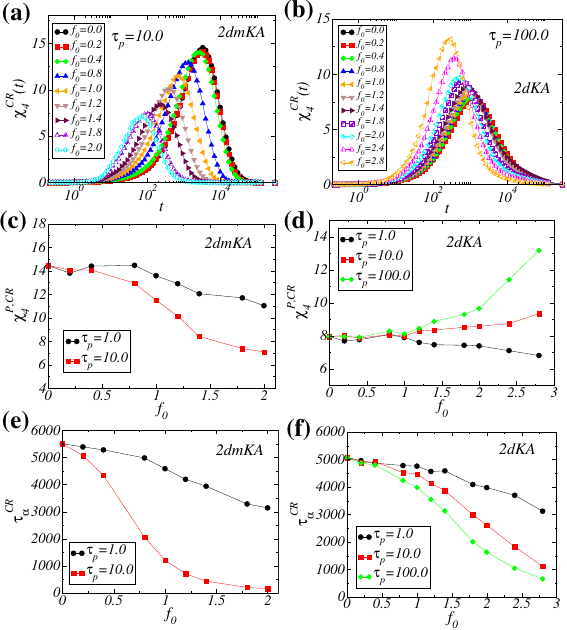}
	\caption{
(a) \& (b) shows the cage-relative $\chi_4(t)$ as a function of time $N = 4000$ with changing $f_0$. (a) is for 2dmKA model with $\tau_p=10$ and (b) is for 2dKA model $\tau_p=100$.  (c) \& (d) shows $\chi^{P,CR}_4$ vs $f_0$. In (c), $\chi^{P,CR}_4$ decreases with increasing activity for the 2dmKA model, but interestingly, in plot (d), $\chi^{P,CR}_4$ shows an increasing trend beyond a certain threshold value of $\tau_p$ for the 2dKA model. (e) \& (f) plots shows $\tau^{CR}_{\alpha}$ vs $f_0$, where in all the cases with increasing activity the $\tau^{CR}_{\alpha}$ decreases.
}
	\label{fig:cageRelCalc}
\end{figure}
\begin{figure*}[!htb] 
	\includegraphics[width=0.98\linewidth]{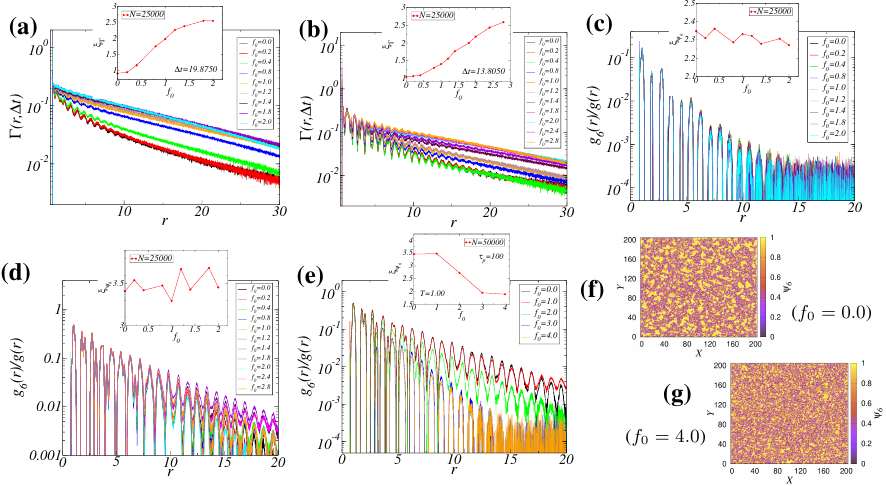}
	\vskip -0.1in
	\caption{{\textbf{Possible Motility Induced Phase Separation (MIPS):}} (a) \& (b) Length scale $\xi_{\Gamma}$ is extracted using excess displacement-displacement correlation ($\Gamma(r,\Delta t)$) calculation, where the $\xi_{\Gamma}$ (Inset) shows increasing nature with activity up to a critical value, for (a) 2dmKA model, for (b) 2dKA model, $N = 25000$. (c) \& (d) Length scale $\xi_{\psi_6}$ (Inset) is extracted using relative hexatic correlation ($g_6(r)/g(r)$), which shows that activity does not lead to enhanced hexatic ordering in the system, (c) \& (d) are for 2dmKA and 2dKA models respectively, $N = 25000$. (e) $g_6(r)/g(r)$ correlation starts to decrease with increasing activity for $\tau_p = 100$ for 2dKA model. Here $\xi_{\psi_6}$ decreases at higher activity $f_0$, $N = 50000$. Hexatic order parameter configuration map for 2dKA model for (f) a passive system ($f_0=0.0$) at $T = 1.0$ and (g) active system with activity $f_0=4.0$, $c = 0.1$, $\tau_p=100.0$, at $T = 1.0$.}
	\label{fig:MIPSCheck}
\end{figure*}

\vskip +0.04in
\noindent{\bf Effect of Local Crystalline Order: }
\SK{In order to investigate the impact of activity on local ordering in disordered solids, we further examined the DH in 2dmKA and 2dKA models by eliminating the effect of long-wavelength phonon modes. We achieved this by computing cage-relative correlation functions, as described in \cite{Mazoyer2009,Illing2017,Vivek2017} and detailed in the SM. By computing $\chi_4^{P,CR}$ and $\tau_\alpha^{CR}$, we were able to isolate the structural part of the relaxation process. Cage-relative correlation functions only consider the displacement of particles relative to their neighbouring particles or cage, thus eliminating the effect of long-wavelength phonon modes in the calculations. In Fig.\ref{fig:cageRelCalc}(A), we present the cage-relative $\chi_4(t)$, or $\chi_4^{CR}(t)$, for the 2dmKA model system with increasing activity while keeping $\tau_p$ and $c$ constant. We observe that the peak height of $\chi_4^{CR}(t)$ systematically decreases with increasing $f_0$.  In panel (C), we plot $\chi_4^{P,CR}$ for the same conditions and note that DH decreases with increasing activity, as previously observed in various 3D systems \cite{Paul2023}. In Fig.\ref{fig:cageRelCalc}(E), we show the variation of $\tau_\alpha^{CR}$ as a function of $f_0$ for two different choices of $\tau_p$. We observe that the relaxation time decreases monotonically with increasing activity. In Fig.\ref{fig:cageRelCalc}(B), we present $\chi_4^{CR}(t)$ for the 2dKA model, which has prominent medium-range crystalline order (MRCO) that increases with increasing supercooling. We observe that the peak height of $\chi_4^{CR}(t)$ increases with increasing activity, while the relaxation time decreases monotonically. This contrasts starkly with the 2dmKA model, which has no prominent local ordering in the studied temperature range. In panel (F), we plot $\tau_\alpha^{CR}$ for three choices of $\tau_p = 1$, $10$, and $100$. In all three cases, we observe that the relaxation time decreases monotonically with increasing $f_0$. However, we note that the variation of $\chi_4^{P,CR}$ is different for these choices. For $\tau_p =1$, the peak height remains nearly constant instead of decreasing, whereas for $\tau_p = 100$, we see a strong increase in peak height as shown in panel (D). These results suggest that activity enhances DH in all active supercooled liquids, but it does so differently for systems with local structural ordering. This is an important finding, as it implies that any observation of increasing $\chi_4(t)$ peak height with decreasing relaxation time in experimental systems in two dimensions might indicate hidden local ordering in the systems. Such systems would behave differently from those that do not tend to grow locally favoured structures (LFS). This observation might have implications for future studies, as it may allow us to quantify local order in various disordered systems and their role in glassy dynamics.
}

\vskip +0.04in
\noindent{\bf Motility Induced Mixing: }
\SK{Next, we focus on the structural analysis of 2dmKA and 2dKA models. First, we compute the displacement-displacement correlation function $\Gamma(r,\Delta t)$ (see SM for details) to evaluate the correlation between particle displacements computed over a time difference of $\Delta t$ near the first peak of $\chi_4(t)$. We present the results of $\Gamma(r,\Delta t)$ for the 2dmKA model in Fig.\ref{fig:MIPSCheck}(A), which shows that the correlation becomes longer range with increasing activity. We also compute the underlying correlation length by integrating the area under the curves. The growth of the correlation length with increasing activity is shown in the inset. Similar results for the 2dKA model system are presented in Fig. \ref{fig:MIPSCheck}(B) and its inset. One can clearly see that for 2dmKA model the correlation length increases first and then saturates to a value at large activity limit. On the other hand, for 2dKA model the correlation length seems to be increasing up to the studied activity window.}

\SK{It is possible that activity can lead to structural ordering in the system, resulting in longer-range spatial correlation, as predicted by the motility-induced phase separation (MIPS) scenario \cite{Cates2015}. However, MIPS typically happens in dilute systems and not in dense limits. To investigate this further, we compute the local hexatic order parameter, $\psi_6$, and the corresponding spatial correlation of $\psi_6$ as $g_6(r) = \langle \psi_6(0) \psi_6(r)\rangle$. In Fig. \ref{fig:MIPSCheck}(C), we show the decay profile of $g_6(r)$ normalized by the radial distribution function $g(r)$. It is evident that the hexatic order does not increase significantly with increasing $f_0$ as the spatial average of $\psi_6(r)$ for the 2dmKA model fluctuates around an average value. Similar results for the 2dKA model are shown in Fig. \ref{fig:MIPSCheck}(D). The average local hexatic order also does not grow beyond its zero activity value. Interestingly, if we increase the persistent time $\tau_p$, we observe that the hexatic order tends to get destroyed systematically, as illustrated in Fig. \ref{fig:MIPSCheck}(e). The inset shows the decrease of mean patch size as computed from the spatial decay of $g_6(r)$ from $\xi_{\psi_6}\sim 3.5$ to $\xi_{\psi_6}\sim 2.0$ with increasing $f_0$ for $\tau_p = 100$. Fig. \ref{fig:MIPSCheck}(f) shows the corresponding snapshots of hexatic field maps for the passive system ($f_0 = 0.0$) and Fig. \ref{fig:MIPSCheck}(g) for the active system with activity $f_0 = 4.0$. Thus, it seems that activity, especially in the large persistence limit, can lead to mixing instead of phase separation in dense systems that have strong tendencies to crystallize. A detailed understanding of this interesting observation warrants further investigations.
}

\begin{figure}[htpb] 
	\includegraphics[width=0.98\linewidth]{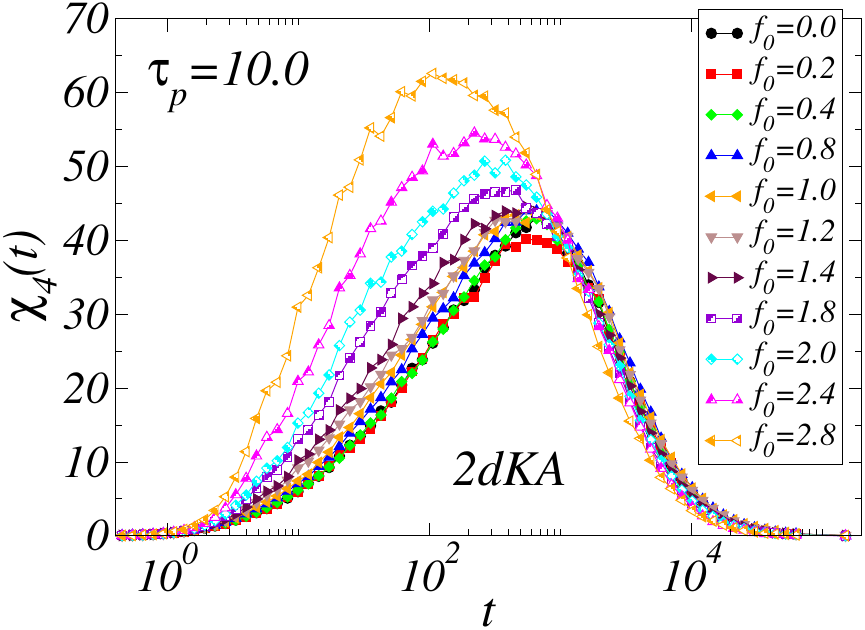}
	\caption{$\chi_4(t)$ vs. $t$ for different $f_0$ of system size $N = 50000$, at fixed $c = 0.1$, $\tau_p=10.0$, using Brownian dynamic simulation for 2dKA model. For better averaging of the $\chi_4(t)$ is calculated using grand-canonical ensemble with the subsystem of size $L/3$ using Block Analysis method (see Block Analysis section in SM).}
	\label{fig:BrownianDyn}
\end{figure}

\noindent {\bf Brownian Dynamics Results:} 
\SK{We performed Brownian Dynamics (BD) simulations to test the robustness of some of our observations, especially that $\chi^{P,CR}_4$ increases with increasing activity despite systematic decrease of relaxation time $\tau^{CR}_{\alpha}$ in the 2dKA model. Further details about Brownian Dynamics simulations can be found in the Methods section. In Fig.\ref{fig:BrownianDyn}, we presented $\chi_4(t)$ as a function of time for various activity strengths. We noticed that although the peak position, which represents the typical relaxation time of the system, decreases with increasing $f_0$, the peak height consistently increases, thus confirming the validity of our MD results. It is important to note that we have only presented results for one type of active driving. It remains unclear whether the observed results are generic across various models of active particles like Active Brownian Particles (ABP), which is a question that needs further exploration. Additionally, we must bear in mind that over-damped Brownian particles are not significantly affected by long-wavelength phonon modes even in passive systems due to the strong suppression of these modes in the presence of the system's internal frictions. However, the observation of MW fluctuations in experimental colloidal systems \cite{Li2019} suggests that activity in these systems will also have stronger MWH fluctuations if measured carefully. 
}	

\vskip +0.02in
\noindent{\bf \large Conclusion:}
\SK{In conclusion, we conducted comprehensive computer simulations to investigate the influence of long-wavelength phonon modes on dynamical heterogeneity, a defining feature of glassy dynamics. Our recent work \cite{Dey2025} demonstrated that long-wavelength phonon modes are amplified by non-equilibrium active forces, resulting in a significant deviation from the predictions of the Mermin-Wagner-Hohenberg (MWH) theorem under equilibrium conditions. The MWH theorem predicts a logarithmic divergence of mean squared positional fluctuations with increasing system size, thereby precluding true long-range positional order in two spatial dimensions (2D). In contrast, the presence of active forces induces a much stronger, power-law divergence. These enhanced fluctuations destabilize active solids even in three dimensions (3D). Our findings underscore the pronounced impact of these long-wavelength fluctuations in two model glasses: one exhibiting local medium-range crystalline order (MRCO) and the other lacking such structural motifs, both in 2D. The influence of active forces is sufficiently strong to generate an additional peak in the four-point susceptibility at short timescales, which can obscure the peak associated with the characteristic $\alpha$-relaxation time at sufficiently high activity and large system sizes. Thus, our results on the effect of MWH fluctuations on dynamical heterogeneity (DH) in the two studied model glasses demonstrate that a straightforward dynamical measurement of four-point susceptibility, $\chi_4(t)$, in the presence of non-equilibrium active forces can yield results that are challenging to interpret.} 

\SK{Furthermore, local structural ordering in these systems exerts significant dynamical influence under active forces even when the effect of the long wavelength modes are systematically taken out. For instance, the 2dKA model, which exhibits local medium-range crystalline order (MRCO), displays an increase in the $\chi_4(t)$ peak with rising activity, despite a systematic decrease in the average characteristic relaxation time. In contrast, the 2dmKA model, lacking predominant local ordering, shows a consistent decrease in the $\chi_4(t)$ peak and a monotonic reduction in relaxation time as activity increases. In these analyses, the influence of long-wavelength phonons has been systematically removed by employing cage-relative correlation functions. These findings may relate to recent observations in Ref.\cite{Cerbino2021} on a confluent monolayer of MCF10A cells, where a sharp increase in the $\chi_4$ peak was reported, even as relaxation time decreased monotonically with increasing activity. This observation suggests that the confluent monolayer may possess local structural ordering while exhibiting glass-like dynamical behavior, similar to the 2dKA model. Further investigation into the interplay between local ordering and activity is warranted.
}

\SK{It is important to highlight that there are systems in which long-range order in 2D can be achieved due to the presence of active noise \cite{Vicsek1995, Toner1998, Palacci2013}, which suppresses the MWH fluctuations unlike our model active system with Run-and-tumble particles (RTPs). In \cite{Galliano2023}, it has been shown that the MWH fluctuation in the system gets suppressed, and the system gets into a long-range ordered state in the presence of a different active noise. This highlights how the behavior of MWH fluctuations in out-of-equilibrium systems is closely tied to the specific nature of the active noise involved. Our own findings \cite{Dey2025} revealed that activity can dramatically enhance the contribution of long-wavelength phonon modes by modifying the effective phonon modes. This opens up an intriguing direction: investigating dynamical heterogeneity in models where active noise suppresses long-wavelength phonon modes, and probing whether local positional order plays a significant role in these systems. In this study, the strength of the activity remained relatively small, but it is known that one finds interesting dynamical behaviour including sub-Arrhenius relaxation dynamics with non-trivial finite size effects \cite{Dey2025_PRE}. Extending this study to high activity limit will be very interesting to see how long-wavelength phonon fluctuations in these extremely activity limits affect the dynamical fluctuations in these 2D systems, especially in the presence of local structural motifs or MRCO.
}

\vskip +0.05in
\noindent{\bf \large Acknowledgements:}
\SK{We acknowledge the funding by intramural funds at TIFR Hyderabad from the Department of Atomic Energy (DAE) under Project Identification No. RTI 4007. SK would like to acknowledge Swarna Jayanti Fellowship Grant Nos. DST/SJF/PSA01/2018-19 and SB/SFJ/2019-20/05 from the Science and Engineering Research Board (SERB) and Department of Science and Technology (DST). SK also acknowledges research support from MATRICES Grant MTR/2023/000079 from SERB.}

\bibliographystyle{unsrt}
\bibliography{AnomalousDH_2D_active.bib}
\end{document}


\title{Supplementary Information : Anomalous Dynamical Heterogeneity in Active Glasses as a Signature of Violation of Mermin-Wagner-Hohenberg Theorem}
\author{Subhodeep Dey}
\author{Smarajit Karmakar}
\email{smarajit@tifrh.res.in}
\affiliation{
Tata Institute of Fundamental Research Hyderabad, 36/P, Gopanpally Village, Serilingampally Mandal, Ranga Reddy District,
Hyderabad, Telangana 500046, India.}

\maketitle

\section{Models and Methods}

In this study, we have used a Binary Mixture of particles, which interact via well-known Lennard-Jones (LJ) potential. This potential has been tuned such that the $2^{nd}$ derivative of the potential is smoothed up to the cut-off distance $r_c$,
\begin{equation}
 \phi(r) = \begin{cases}
 4\epsilon_{\alpha\beta}\left[\left(\frac{\sigma_{\alpha\beta}}{r}\right)^{12} -\left(\frac{\sigma_{\alpha\beta}}{r}\right)^{6} + c_0 + c_2r^2 \right] &, r < r_c \\
 0 &, r \geq r_c
 \end{cases}
\end{equation}
Here, $\{\alpha,\beta\}$ refers to the type of the particles, which can be A (large) or B (small) type. We studied two types of glass-forming liquids; one is 2D modified Kob-Andersen (2dmKA), where A:B number ratio is $65:35$, and other one is 2D Kob-Andersen model (2dKA), where A:B number ratio is $80:20$ \cite{Tah2018}. The interaction strengths are $\epsilon_{AA}=1.0$, $\epsilon_{AB}=1.5$, $\epsilon_{BB}=0.5$ and the diameter of the particles are $\sigma_{AA}=1.0$, $\sigma_{AB}=0.8$, $\sigma_{BB}=0.88$. The interaction cut-off is $r_c=2.5\sigma_{AB}$. Here the reduced unit of length, energy and time are given by $\sigma_{AA}$, $\epsilon_{AA}$ and $\sqrt{\frac{\sigma^2_{AA}}{\epsilon_{AA}}}$ respectively. For all cases, the system's particle density ($\rho$) is fixed at $1.2$, and the integration step ($\delta t$) is set at $0.005$. For 4D Kob-Andersen model (4dKA), A:B number ratio is $80:20$ with density ($\rho$) is $1.474$ \cite{Eaves2009}.

\section{Activity in the system: Run and Tumble particle (RTP) model}
To introduce activity in the system, we have given additional active force, $f_0$, to a set of selected (tagged) active particles in addition to their mutual particle-particle interaction coming from the inter-particle potential. Here, we have selected fraction ($c$) of active particles in the system. We assign each active particle a active direction ($k_x,k_y$) randomly in a manner that the total active force applied to all the active particles sum up to zero. This done to maintain the momentum conservation in the system. The persistent time $\tau_p$ is the time until which the active force on each particle acts in the same direction and after $\tau_p$, the directions are randomly reshuffled. The active force on the $i^{th}$ particle in 2D can be written as, 
\begin{equation}
 \vec{F}^A_i = f_0 (k^i_x \hat{x}+k^i_y \hat{y}),
\end{equation}
where the active direction (kx, ky) is chosen from $\pm1$, and the number of total active particles is taken to be an even number to maintain the total active momentum to be zero along all directions, i.e., $\sum_{\alpha,i} k^i_{\alpha}=0$. Now, in this active system, there are three tuning parameters $c$, $f_0$ and $\tau_p$. We have different activity parameters for 2dmKA and 2dKA systems depending on the activity limit. To start with, we first fixed $\tau_p = 1.0$ for all the cases, then we changed the $f_0$ and $c$ separately. For 2dmKA, we have varied $f_0$ $\in$ [0.0-2.0], and for 2dKA, we have varied $f_0$ $\in$ [0.0-2.8], keeping $c$ fixed at $0.1$ for both cases. Again, for both 2dmKA and 2dKA, while varying c $\in$ [0.1-0.6], we kept $f_0$ fixed at $1.0$. Later, to study the effect of persistent time, we have varied $f_0$ at $\tau_p = 10$ and $100$ at fixed $c = 0.1$.

\section{Thermostat}
In this study, we have used the three-chain Nos\'e-Hoover thermostat \cite{Allen} to get the required temperature. This thermostat is known to give a true canonical ensemble in an equilibrium system, which is also suitable for a canonical ensemble in an out-of-equilibrium system. Here, we have taken the coupling relaxation time ($\tau_T$) of the thermostat $10$ times the integration time step ($\delta t$).
  
\section{Two-point correlation Function, $Q(t)$}
To compute two-point density-density correlation, we have considered a simpler form of overlap correlation function $Q(t)$ defined as 
\begin{equation}
 Q(t) = \frac{1}{N} \sum^N_{i=1} w(|\vec{r}_i(t)-\vec{r}_i(0)|),
\end{equation}
where $w(x)$ is a window function, which is $1$ for $x<a$ and $0$ otherwise. $\vec{r}_i(t)$ is the position of the $i^{th}$ at time $t$. Here, parameter `$a$' is chosen to remove the vibration due to the caging effect. One can choose the value of 'a' from the plateau of a `mean-square displacement' (MSD) in the supercooled temperature regime. For all cases, the value of $a$ is set to $0.3$. From this two-point correlation function, we define a relaxation time $\tau_{\alpha}$, as $\left<Q(t=\tau_{\alpha})\right>=1/e$, where $\left<...\right>$ is ensemble average and $e$ is the base of the natural logarithm. The system is equilibrated for $150\tau_{\alpha}$ and then we ran the simulations for another $150\tau_{\alpha}$ for measuring various dynamical and thermodynamic quantities. We chose $\tau_{\alpha} \sim 2000$. For system size $N \leq 10000$, we have averaged our data over $32$ statistically independent ensembles runs, and for system size $N > 10000$, we have taken $6$ ensembles for averaging.

\section{Four-point correlation Function, $\chi_4(t)$}  
The four-point correlation function can be measured from the fluctuation of the two-point correlation function Q(t) as,
\begin{equation}
 \chi_4(t) = N \left[\left<Q(t)^2\right> - \left<Q(t)\right>^2\right],
\end{equation}
where $\left<...\right>$ is the ensemble average. $\chi_4 (t)$ is a well-known quantifier of dynamic heterogeneity (DH) in the system. Dynamic heterogeneity broadly refers to the heterogeneous dynamical relaxation processes in various parts of the system. This happens due to different population of slow and fast-moving particles in the system. DH reaches its maximum round the relaxation time $\tau_{\alpha}$ and the peak is defined as $\chi_4(t=\tau_{\alpha}) \simeq \chi^P_4$. In this study, due to the activity and increasing system size, one observes the emergence of an additional peak at the early-beta regime in $\chi_4(t)$. We focus on the first peak of the $\chi_4(t)$ in the early-beta regime and define the first peak as $\chi_4(t=t^*) \simeq \chi^{P1}_4$, where $t^*$ is the characteristic time at which the first peak appears.

\section{Cage-relative displacement}
At large system size limit with increasing activity, the system starts to show increasing fluctuations at the short time of $\chi_4(t)$. The fluctuations becomes so large that it nearly mask the heterogeneity peak $\chi^P_4$ around relaxation time $\tau_{\alpha}$. To separate the influence of the collective behaviour affecting the system dynamics due to glassy relaxation process and not from the long wavelength phonon fluctuations, we have computed the cage-relative (CR) displacement \cite{Mazoyer2009, Vivek2017, Illing2017} of the individual particle $i$, such as 
\begin{equation}
 \vec{r}_{i,CR}(t) = \left[ \vec{r}_i(t) - (\vec{r}_{i,nn}(t) - \vec{r}_{i,nn}(0) ) \right],
\end{equation}
where $\vec{r}_{i,nn}(t)$ is the position of the center of mass of $N_{nn}$ nearest neighbours of $i^{th}$ particles. Again it is defined as
\begin{equation}
 \vec{r}_{i,nn}(t) = \frac{1}{N_{nn}} \sum^{N_{nn}}_{j=1} \left[ \vec{r}_j(t) - \vec{r}_j(0) \right].
\end{equation}
Here, we have used cut-off value $r^{nn}_c=1.3$ for the first nearest neighbours $N_{nn}$. After identifying the $N_{nn}$ particles at the initial time or at each time origin, we track these particles' ($N_{nn}$) dynamics at the later time for the computation of the cage relative correlation functions.  We use this cage-relative displacement to calculate $Q(t)$, $\chi_4(t)$ and MSD, respectively.

\section{choice of temperature for different activity}
\begin{figure*}[!htpb]
\centering
\includegraphics[width=0.82\textwidth]{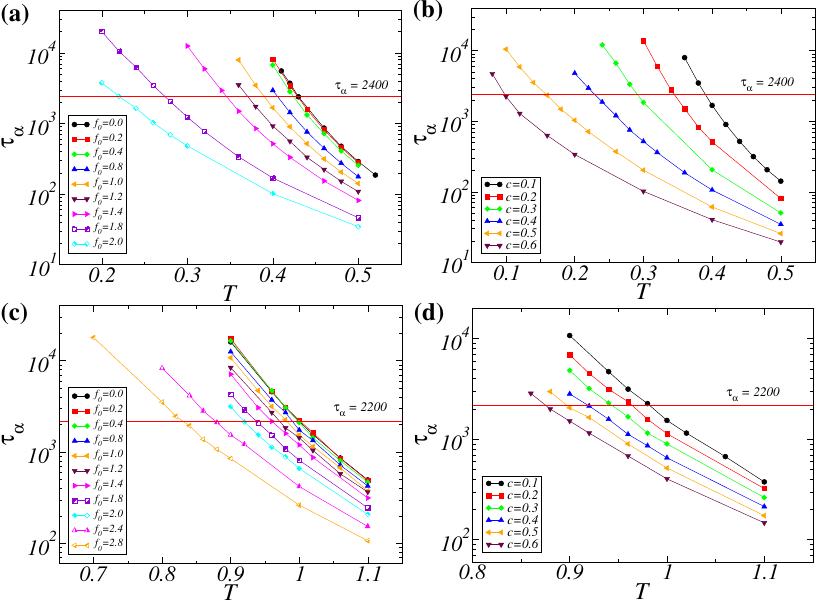}
\caption{Left panel: (a) \& (c) plots represents $\tau_{\alpha}$ vs T for different $f_0$ and fixed c=0.1 and $\tau_p=1.0$. Right panel: (b) \& (d) plots  represents $\tau_{\alpha}$ vs T for different c and fixed $f_0=1.0$ and $\tau_p=1.0$ . Top panel: (a), (b) plots are for the 2dmKA model and bottom panel: (c), (d) plots are for the 2dKA model.}
\label{fig:tauAlphavsT_C_f0_2dmKA_2dKA_merge}
\end{figure*}
\begin{figure*}[!htpb]
\centering
\includegraphics[width=0.82\textwidth]{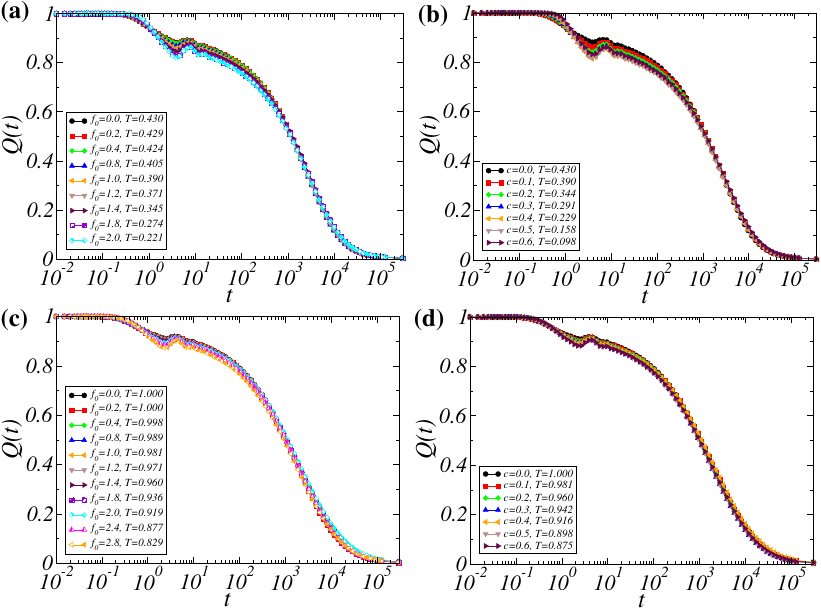}
\caption{Left panel: (a) \& (c) plots represent two-point correlation vs for different $f_0$ at fixed c=0.1, $\tau_p=1.0$ gives similar relaxation time. Right panel: (b) \& (d) plots represent two-point correlation vs for different c at fixed $f_0=1.0$, $\tau_p = 1.0$ gives similar relaxation time. Top panel: (a), (b) plots are for the 2dmKA model and bottom panel: (c), (d) plots are for the 2dKA model.}
\label{fig:relax_N_1e3_C_f0_2dmKA_2dKA_merge}
\end{figure*}

 
As activity in each model (2dmKA and 2dKA) can be tuned either by tuning the fraction of active particles ($c$) or the active force ($f_0$) or the persistent time ($\tau_p$), it is important to identify a correct measure to compare these systems across the parameter space. For this we choose relaxation time $\tau_{\alpha}$ is the system's characteristics for comparison across various values of $c$, $f_0$ and $\tau_p$. For a given set of $c$, $f_0$ and $\tau_p$, if we vary $T$, then the obtained relaxation time is found to obey the well-known `Vogel-Fulcher-Tammann' (VFT) relation, defined as
\begin{equation}
 \tau_{\alpha} = \tau_0 \exp[A/T-T_0].
\end{equation}
We then take an intermediate system size and fix the $\tau_{\alpha}$ for different activities at the suitable temperature using the VFT relation. For 2dmKA, we have taken the relaxation time $\tau_{\alpha} \simeq 2400$ which happens at $T = 0.43$ of a passive system with $N = 1000$. We then fix these temperatures for all other system sizes for a direct comparison. For 2dmKA model system, we firstly change the $f_0$ from $0.0$ to $2.0$ by fixing $c = 0.1$ and $\tau_p = 1.0$. Next, we have changed the $c$ from $0.0$ to $0.6$ by fixing $f_0 = 1.0$ and $\tau_p = 1.0$. Similarly, for 2dKA, we have taken the relaxation time $\tau_{\alpha} \simeq 2200$ at $T = 1.00$ of a passive system with $N = 1000$ and fixed it for all other activities and system sizes. We changed the values of $f_0$ from $0.0$ to $2.8$ by fixing $c = 0.1$ with $\tau_p = 1.0$. Then, we change the $c$ from $0.0$ to $0.6$ by fixing $f_0 = 1.0$ and $\tau_p = 1.0$. From Fig. \ref{fig:tauAlphavsT_C_f0_2dmKA_2dKA_merge}, we have used the temperature for the respective activity parameter and fixed it for the rest of the study. Again, in Fig. \ref{fig:relax_N_1e3_C_f0_2dmKA_2dKA_merge}, we have shown that the choice of temperature corresponding to different activity is giving similar relaxation time for the system size $N = 1000$.

\section{large system size in high activity limit}
\begin{figure}[!htpb]
\centering
\resizebox{80mm}{60mm}{\includegraphics{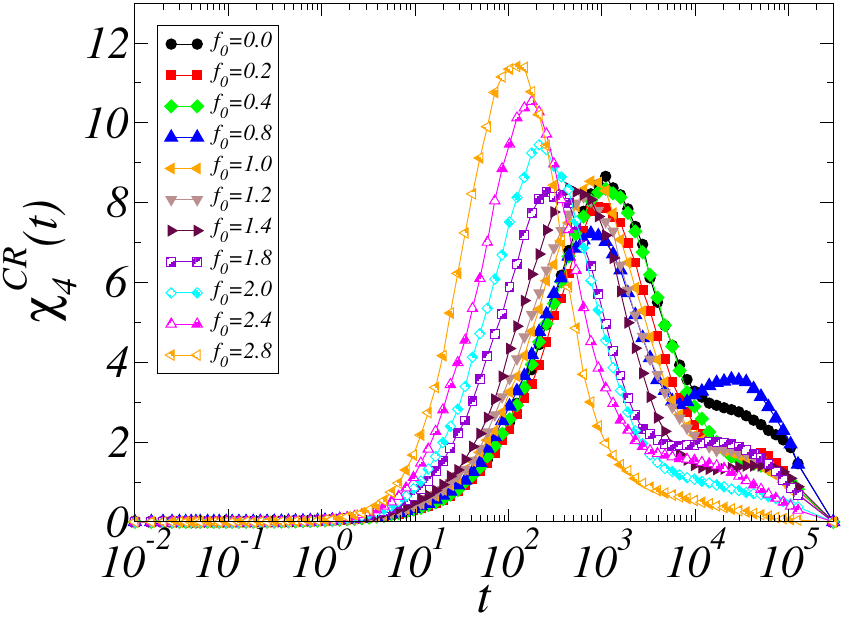}}
\caption{($\chi^{CR}_4(t)$ vs. t) for different $f_0$ of system size N=50000, at fixed c=0.1, $\tau_p=10.0$, 2dKA.}
\label{fig:2dKA_N_5e4_1e1_CM_chi4_f0}
\end{figure}

\begin{figure}[!htpb]
\centering
\resizebox{80mm}{60mm}{\includegraphics{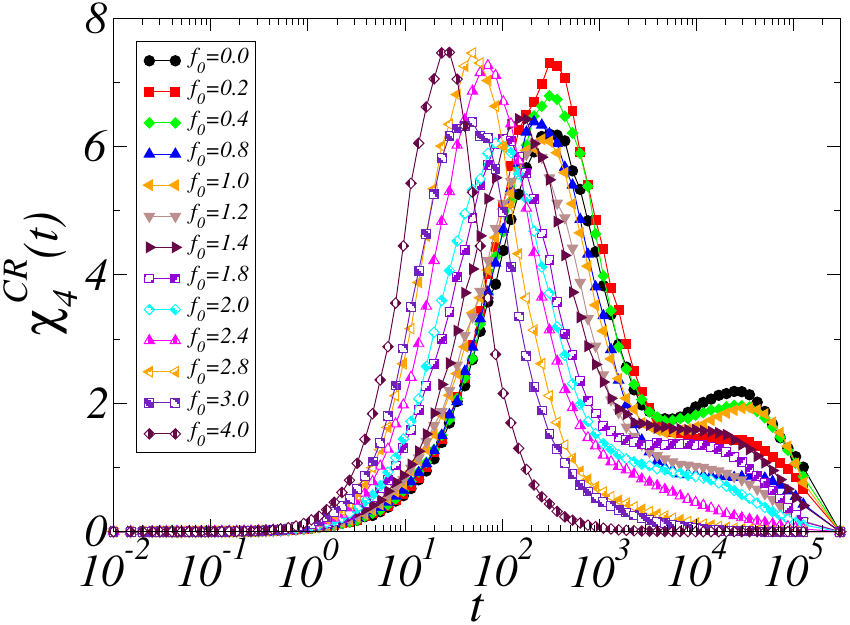}}
\caption{($\chi^{CR}_4(t)$ vs. t) for different $f_0$ of system size N=50000, at fixed c=0.1, $\tau_p=10.0$ and T=1.100, 2dKA.}
\label{fig:2dKA_N_5e4_1e1_CM_chi4_f0_T_1.100}
\end{figure}

To understand the effect of changing $f_0$ in the high persistent time limit ($\tau_\alpha = 10$ and $100$), we have taken large system $N = 50000$  for 2dKA model simulations. Fig. \ref{fig:2dKA_N_5e4_1e1_CM_chi4_f0} shows that the $\chi^{P,CR}_4$ starts to increase from activity $f_0=2.0$, where at small activity limit $\chi^{P,CR}_4$ fluctuates around 7 and 8. To see the effect of activity alone, we have fixed the system as high-temperature T=1.100 for all activity. From Fig. \ref{fig:2dKA_N_5e4_1e1_CM_chi4_f0_T_1.100} we observe that there is no clear trend of increasing $\chi^{P, CR}_4$ with increasing activity, it fluctuates around 6 to 8 for different activity. Both observations show that $\chi^{P, CR}_4$ does not decrease with increasing activity even at very low $\tau_{\alpha}$. We observe an additional peak arising for small activity at a long time limit; this effect could be due to less averaging at a longer time.

\begin{figure}[!htpb]
\centering
\resizebox{80mm}{60mm}{\includegraphics{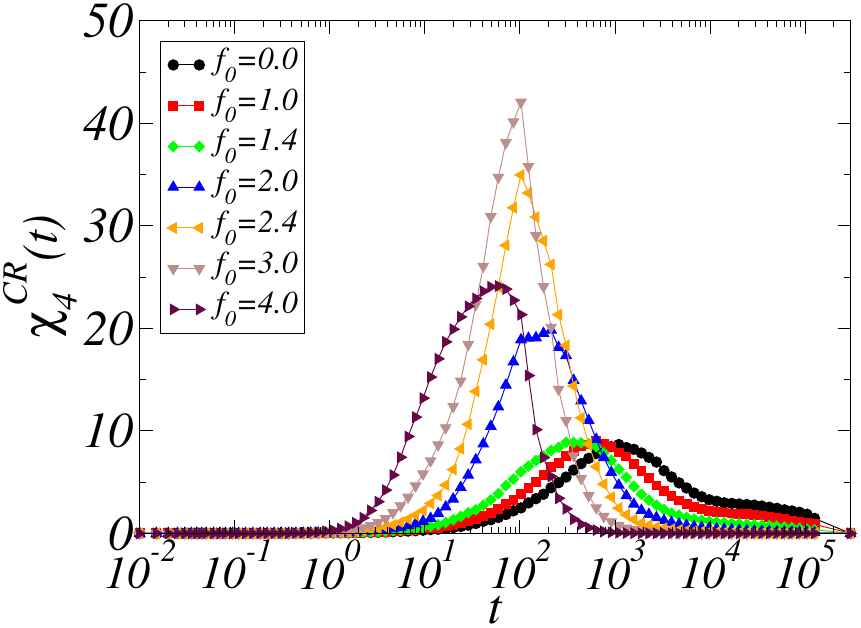}}
\caption{($\chi^{CR}_4(t)$ vs. t) for different $f_0$ of system size N=50000, at fixed c=0.1, $\tau_p=100.0$ and T=1.000, 2dKA.}
\label{fig:2dKA_N_5e4_1e2_CM_chi4_f0_T_1.000}
\end{figure}

\section{Block Analysis}
To study the effect of other fluctuations including number, density, composition, temperature and activity fluctuations, we used the method of Block Analysis as detailed in Ref.\cite{Chakrabarty2017}, where one divides the whole system into multiple subsystems of equal linear sizes and then computes the relevant correlation functions. This method is very useful for colloidal experiments, in which often one records the data over a small measurement region in the whole sample. The two-point correlation for a single subsystem can be defined as,
\begin{equation}
 Q(L_B,t) = \frac{1}{n_i} \sum^{n_i}_{j=1} [w(r_j(t)-r_j(0))],
\end{equation}
where $L_B=(N/N_B)^{1/3}$ is the subsystem size, $N_B$ is the number of subsystems (blocks) and $n_i$ is the number of particles in the $i^{th}$ block. The average two-point correlation function is given by,
\begin{equation}
 \left<Q(L_B,t)\right> = \frac{1}{N_B} \sum^{N_B}_{i=1} Q(L_B,t).
\end{equation}
Similarly, the four-point dynamic susceptibility for each block can be computed as,
\begin{equation}
 \chi'_4(L_B,t) = \left[\left<Q(L_B,t)^2\right> - \left<Q(L_B,t)\right>^2\right].
\end{equation}
The system averaged four-point dynamic susceptibility of the sub-system ($L_B$) is given by,
\begin{equation}
 \chi_4(L_B,t) = \left(\frac{N}{N_B}\right) [\chi'_4(L_B,t)].
\end{equation}
  Here $\left<...\right>$ denotes the average over different grand-canonical ensembles, i.e., for different time origins and sub-systems.

\begin{figure}[!htpb]
\centering
\resizebox{80mm}{60mm}{\includegraphics{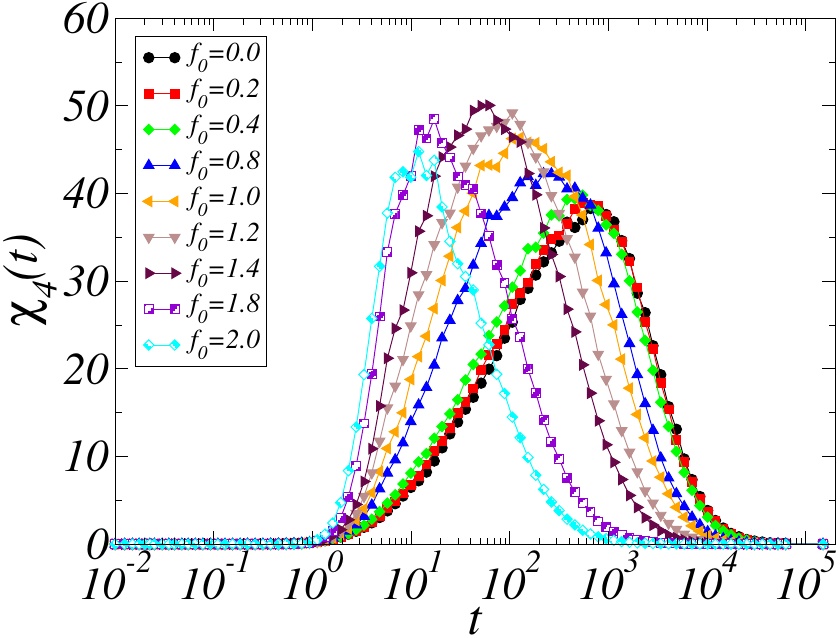}}
\caption{$\chi_4(t)$ vs. t for different $f_0$ of system size $N = 50000$, $n_B$=3 at fixed $c=0.1$, $\tau_p=10.0$, using Brownian dynamic simulation, 2dmKA.}
\label{fig:2dmKA_BD_N_5e4_1e1_nB_3_chi4_f0}
\end{figure}
\section{Brownian dynamics}
\begin{figure}[!htpb]
\centering
\resizebox{80mm}{55mm}{\includegraphics{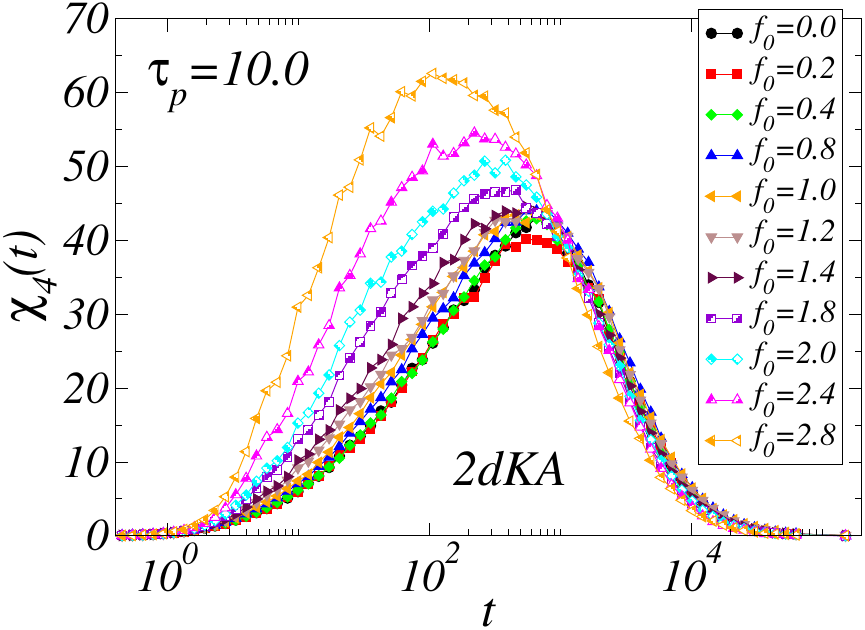}}
\caption{$\chi_4(t)$ vs. t for different $f_0$ of system size $N=50000$, $n_B =3$ at fixed $c=0.1$, $\tau_p=10.0$, using Brownian dynamic simulation, 2dKA.}
\label{fig:2dKA_BD_N_5e4_1e1_nB_3_chi4_f0}
\end{figure}

\begin{figure}[!htpb]
\centering
\vskip -0.1in
\resizebox{80mm}{55mm}{\includegraphics{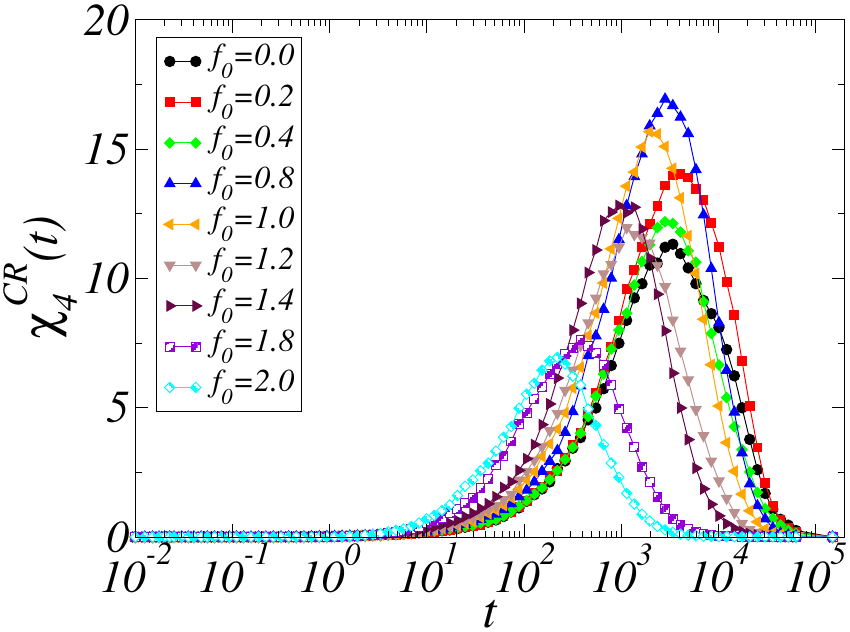}}
\caption{($\chi^{CR}_4(t)$ vs. t) for different $f_0$ of system size N=50000, at fixed c=0.1, $\tau_p=10.0$, using Brownian dynamic simulation, 2dmKA.}
\label{fig:2dmKA_BD_N_5e4_1e1_CM_chi4_f0}
\end{figure}

\begin{figure}[!htpb]
\centering
\resizebox{80mm}{55mm}{\includegraphics{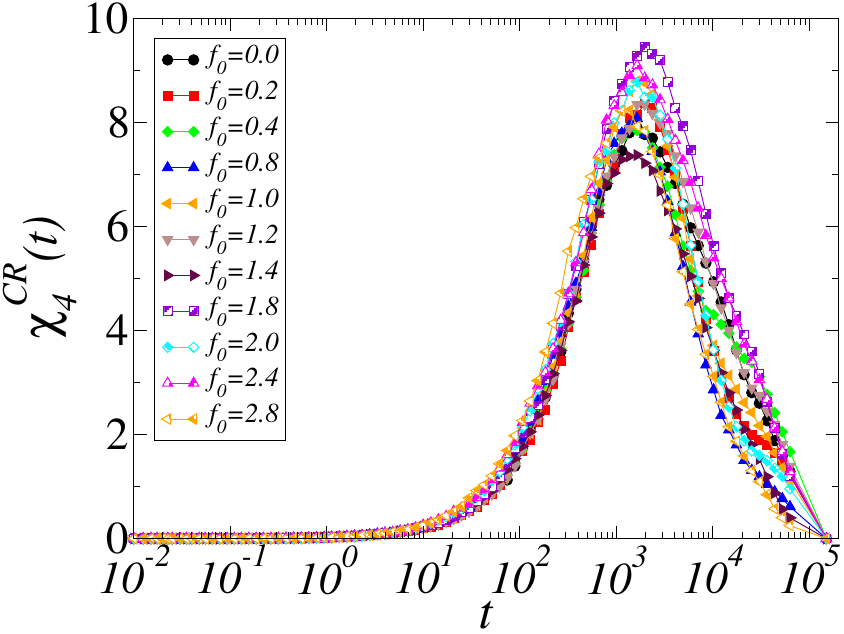}}
\caption{($\chi^{CR}_4(t)$ vs. t) for different $f_0$ of system size N=50000, at fixed c=0.1, $\tau_p=10.0$, using Brownian dynamic simulation, 2dKA.}
\label{fig:2dKA_BD_N_5e4_1e1_CM_chi4_f0}
\end{figure}

We have also performed Brownian dynamics (BD) simulations \cite{Beard2000} to understand the effect of activity in over-damped limits. We have set the system's diffusivity ($D_0$) to $1.0$. As we know, the long-time behaviour of both Brownian dynamics (BD) and molecular dynamics (MD) is the same; we have kept all the parameters in the BD simulation the same as in the case of the MD simulation. Now, we have taken a single canonical ensemble of extensive system $N = 50000$. Then, we used the Block Analysis to get the grand canonical ensemble average of the dynamical susceptibility ($\chi_4(t)$). We have taken a subsystem of size $L_B = L/3$, giving us $8$ subsystems for averaging. As BD is in the over-damped limit here, the elastic long-wavelength mode should be suppressed considerably, and it should not mask the $\chi_4$ peak around $\tau_{\alpha}$ like MD simulation. In Fig. \ref{fig:2dmKA_BD_N_5e4_1e1_nB_3_chi4_f0} $\chi^p_4$ fluctuates around 40 with changing activity in 2dmKA system. From Fig. \ref{fig:2dKA_BD_N_5e4_1e1_nB_3_chi4_f0}, we can observe that $\chi^P_4$ is increasing with increasing activity from $f_0 = 1.8$. Again, we have measured the cage-relative $\chi_4(t)$ to neglect any possible collective motion in the system. In Fig. \ref{fig:2dmKA_BD_N_5e4_1e1_CM_chi4_f0} for 2dmKA $\chi^{P, CR}_4$ tends to go down at higher activity, but In Fig. \ref{fig:2dKA_BD_N_5e4_1e1_CM_chi4_f0} we can observe the $\chi^{P, CR}_4$ to increase for different activity for 2dKA, unlike the normal glass (2dmKA). This analysis has been done on a single ensemble of $N = 50000$ system size; here also one observes the long time peak in Fig. \ref{fig:2dKA_BD_N_5e4_1e1_CM_chi4_f0} for 2dKA, now to understand this effect one needs further studies.

\newpage

 \section{Displacement-displacement correlation function, $\Gamma(r,\Delta t)$}
The dynamical length scale of the system $\xi_d$ can be computed independently 
by computing the displacement-displacement correlation function $g^{uu}(r,t^*)$ at 
$t^*$ \cite{TahPRR, Poole1998}. It is defined as,
\begin{equation}
g^{uu}(r,\Delta t)=\frac{\left\langle  \sum\limits_{i,j=1,j\neq i}^Nu_i(0,\Delta t)u_j(0,\Delta t)\delta(r-|\textbf{r}_{ij}(0)|)\right\rangle}{4\pi r^2\Delta rN\rho \langle u(\Delta t)\rangle^2}
\label{guu}
\end{equation}  
 where, $u_i(t,\Delta t)=|\textbf{r}_i(t+\Delta t)-\textbf{r}_i(t)|$, and $\langle u^2(\Delta t)\rangle=\langle\frac{1}{N}\sum_{i=1}^N
 \textbf{u}_i(t,\Delta t).\textbf{u}_i(t,\Delta t)\rangle$. $g^{uu}(r,\Delta t)$ is calculated at time $\Delta t=t^*$, along with the usual pair 
 correlation function $g(r)$ defined as,
 \begin{equation}
 g(r)=\frac{\left\langle\sum\limits_{i,j=1,j\neq i}^N\delta(r-|\textbf{r}_{ij}(0)|) \right \rangle} {4\pi r^2\Delta r N\rho}
\end{equation}  
 
For far enough particles, the displacement over a large enough time duration would be decorrelated, and $g^{uu}$ would equal g(r). So the quantity $\Gamma(r,\Delta t) = g^{uu}(r,\Delta t)/g(r)-1.0$ would decay to zero as a function of $r$. If one assumes the decay to be exponential, the area under the curve will provide us with the correlation length.

\section{Orientational Order}

Per particle hexatic order parameter:
 \begin{equation}
 \psi_6(\vec{r_i},t)=\frac{1}{n_i} \sum^N_{j\in n_i} exp(i6\theta_{ij}),
\end{equation}
where $n_i$ is the number of nearest neighbours of particles $i$. 
Hexatic correlation function:
 \begin{equation}
 g_6(r) = \frac{1}{\rho N} \left<\sum_{i\neq j} \psi_6(\vec{r_i}) \psi^*_6(\vec{r_j}))\delta(|\vec{r}-\vec{r_j}+\vec{r_i}|)\right>
\end{equation}
This Hexatic correlation function has been normalised using pair correlation function g(r).

\section{Additional Relaxation Time data}

\begin{figure*}[!htpb]
\centering
\includegraphics[width=0.90\textwidth]{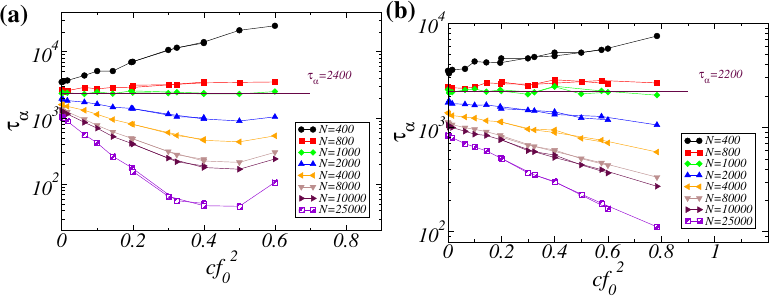}
\caption{(a) \& (b) plots represents $\tau_{\alpha}$ vs $cf^2_0$. Here, both plots show the change in $\tau_{\alpha}$ for different activity at their respective temperature, and the temperature is chosen such that $\tau_{\alpha}$ for the N=1000 system remains. $\tau_{\alpha}$ is set at 2400 for 2dmKA, and $\tau_{\alpha}$ is set at 2200 for 2dKA. Temperature is kept the same for all the respective system sizes if not mentioned specifically. (a) plot is for the 2dmKA model, and (b) plot is for the 2dKA model.}
\label{fig:tauAlphavscf02_N_2dmKA_2dKA_merge}
\end{figure*}

In Fig. \ref{fig:tauAlphavscf02_N_2dmKA_2dKA_merge} its shows the relation of $\tau_{\alpha}$ with $cf^2_0$ for both 2dmKA and 2dKA. One striking observation of the 2dmKA is the non-monotonous nature of the $\tau_{\alpha}$ with $cf^2_0$, which is not present for the case of 2dKA in the present system.

\section{Cage-relative Relaxation Time}

\begin{figure*}[!htpb]
\centering
\includegraphics[width=0.90\textwidth]{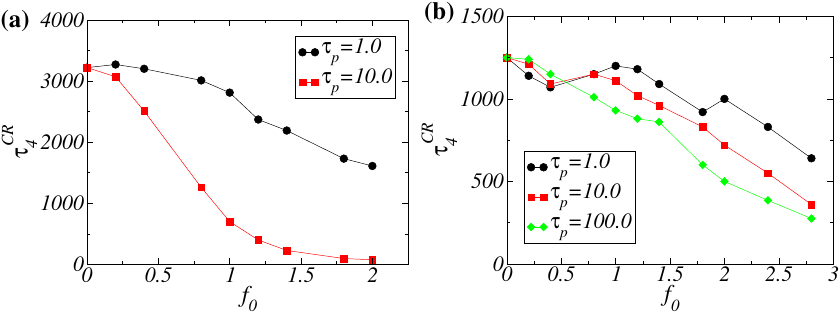}
\caption{Peak time of $\chi_4^{CR}$, $\tau_4^{CR}$ for increasing activity $f_0$ for different persistent $\tau_p$. (a) For normal disorder system (2dmKA) $\tau_4^{CR}$ decreases systematically for increasing $f_0$ for studied $\tau_p$. (b) For medium range order glassy system (2dKA) $\tau_4^{CR}$ decreases systematically for increasing activity $f_0$ in the limit $f_0>1.0$ for different persistent time $\tau_p$.}
\label{fig:tau0p_f0_CM_taup}
\end{figure*}

In this activity regime, the cage-relative motion of the system shows a non-trivial increase in dynamical heterogeneity near the slow alpha-relaxation time. In Fig. \ref{fig:tau0p_f0_CM_taup} (a), we have shown that the peak time of $\chi_4^{CR}$ given by $\tau_4^{CR}$ decreases rapidly with increasing $\tau_p$ for a normal disorder system (2dmKA). Also in Fig. \ref{fig:tau0p_f0_CM_taup} (b), we have shown that for medium range order in the glassy system (2dKA) $\tau_4^{CR}$ decreases with increasing $\tau_p$ when the activity $f_0$ is more than 1.0. Here, the temperature of the system is chosen to keep the system in the supercooled regime for the system size N=1000. So the non-monotonic increase in the small $f_0$ limit could be reminiscent of the non-trivial system size effect of the $\tau_p$. Again, we have also pointed out similar results using the scaling description study of the high-activity glassy system (in Chapter 5).

\section{Displacement field}

\begin{figure*}[!htpb]
\centering
\includegraphics[width=0.90\textwidth]{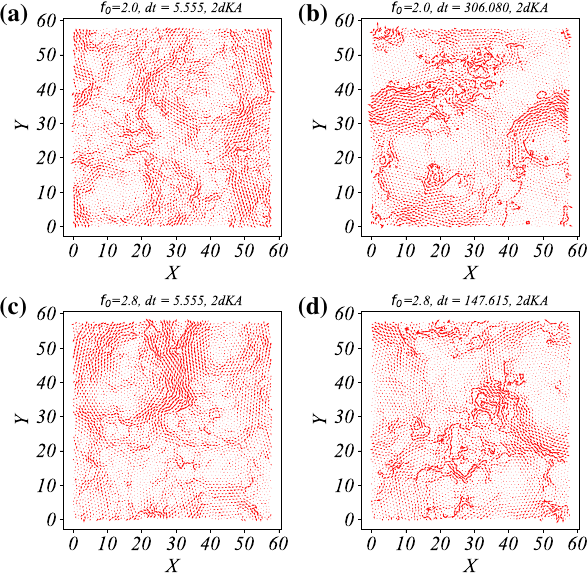}
\caption{Displacement field for 2dKA system with fixed persistent time $\tau_p=10.0$ for intermediate activity $f_0=2.0$ (a, b) and larger activity $f_0=2.8$ (c, d). For activity $f_0=2.0$, the short time dt=5.555 (a) shows a larger domain of collective motion than the long time dt=306.080 (b). Similarly, for $f_0=2.8$, short time limit dt=5.555 (c) shows a larger domain of collective motion than long time dt=147.615 (d). Here, short time dt is approximately the time at which the first peak of the dynamical heterogeneity is observed, and long time dt is approximately the time at which the long-time dynamical heterogeneity peak is observed for the system size N=4000.}
\label{fig:DH_2dKA_taup_1e1}
\end{figure*}

To better understand the heterogeneous motion of the fast-moving particles in the system, we have shown the displacement field for time $dt$ for the 2dKA system of system size N=4000, as shown in Fig. \ref{fig:DH_2dKA_taup_1e1}. Here the persistent time $\tau_p=10.0$ is fixed for $f_0$ 2.0 (a, b) and 2.8 (c, d) for short time (a, c) and long time (b, d) dynamics. Here, short time dt is approximately the time at which the first peak of the dynamical heterogeneity is observed, and long time dt is approximately the time at which the long-time dynamical heterogeneity peak is observed for system size N=4000. In Fig. \ref{fig:DH_2dKA_taup_1e1}, it is shown that the short-time (a, c) displacement has a larger domain of collective motion than the long-time (b, d) displacement.

\bibliographystyle{apsrev4-2}
\bibliography{AnomalousDH_2D_Supple}